\documentclass[aps,prd,preprint,floatfix,superscriptaddress,nofootinbib,longbibliography,a4paper]{revtex4-1}
\pdfoutput=1
\usepackage{color}
\usepackage{graphicx}
\usepackage{textcomp}
\usepackage{subfigure}
\usepackage{feynmp}
\usepackage{amsmath,amssymb, array}
\usepackage{enumerate}
\usepackage{slashed}
\usepackage[normalem]{ulem}
\usepackage{hhline}
\usepackage{hyperref}
\usepackage{empheq}
\usepackage{cases}
\usepackage{longtable}
\hypersetup{colorlinks=true,
    citecolor=blue,
	linkcolor=blue,
	filecolor=magenta,      
	urlcolor=blue}
\newcommand{\mathsym}[1]{{}}

\newcommand{\beqa}{\begin{eqnarray}}
\newcommand{\eeqa}{\end{eqnarray}}
\newcommand{\be}{\begin{equation}}
\newcommand{\ee}{\end{equation}}
\newcommand{\ba}{\begin{array}} 
\newcommand{\ea}{\end{array}}
\newcommand{\taub}{\overline{\tau}}

\newcolumntype{C}{>{$}c<{$}}
\def\vev#1{\langle #1\rangle}

\begin{document} 

\title{Restoring minimal modular neutrino masses via K\"ahler $R$-parity violation}
\bigskip
\author{Anjan S. Joshipura}
\email{anjanjoshipura@gmail.com}
\affiliation{Theoretical Physics Division, Physical Research Laboratory, Navarangpura, Ahmedabad-380009, India}
\author{Ketan M. Patel}
\email{ketan.hep@gmail.com}
\affiliation{Theoretical Physics Division, Physical Research Laboratory, Navarangpura, Ahmedabad-380009, India}

\begin{abstract}
The possibility of describing neutrino masses through a single multiplet of modular forms in a modular symmetric framework is reanalysed. The minimal model achieving this is Feruglio's model based on modular $A_4$ symmetry, which contains a chiral flavon triplet $\phi$, with all neutrino masses and mixing determined by a superfield triplet of modular forms $Y(\tau)$. We point out that, in local supersymmetry, these superfields by themselves allow an additional $R$-parity-violating contribution to neutrino masses originating from the K\"ahler potential of the model. The $R$ violation manifests through a bilinear term $\epsilon_i L_i H_u$ in the effective superpotential, with parameters $\epsilon_i$ determined in terms of $Y(\tau)$ and its derivatives. The K\"ahler potential also mixes neutrinos with the fermionic component of $\phi$, leading to seesaw neutrino masses when these fermions are massive. Two separate cases are analysed, with neutrinos obtaining mass through the Weinberg operator or through the $\phi$-induced seesaw. Unlike the original model, adding the $R$-violating contribution allows accurate reproduction of neutrino mixing angles and mass ratios for both the normal and inverted orderings in either case, with the Weinberg operator predicting a sum of neutrino masses close to the cosmological limit and the seesaw mechanism predicting ones comfortably below it.
\end{abstract}

\maketitle

\section{Introduction}
\label{sec:intro}
Modular symmetry arising in string compactification has played an important role in the bottom-up approach \cite{Feruglio:2017spp} to understanding the structure of the observed fermion masses and mixing angles; for recent reviews, see \cite{Feruglio:2019ybq,Ding:2023htn,Kobayashi:2023zzc}. The modular symmetric framework is described in the context of supersymmetry by three elements: (i) the underlying symmetry group $\Gamma_N$, which is a finite discrete modular group; (ii) the modulus $\tau$, which is a chiral superfield; and (iii) modular forms $Y_r^{(k_y)}(\tau)$, which transform as an $r$-dimensional representation of the underlying $\Gamma_N$ and possess modular weight $k_y$. Given the group $\Gamma_N$, the modular forms are completely determined as functions of $\tau$ alone. The scalar components of $Y_r^{(k_y)}(\tau)$ play the role of the Yukawa couplings. This aspect distinguishes modular symmetry from conventional discrete symmetries, in which the Yukawa couplings are functions of the unknown vacuum expectation values of the flavon fields that break the underlying symmetry group. However, in the modular symmetric framework, the arbitrariness enters through the choice of the representations and modular weights of the various matter fields, and it is these choices that determine the structure of the effective Yukawa couplings and, ultimately, the spectrum of the matter fields and their interactions. These choices are not unique and can lead to a large number of possibilities in determining the underlying model \cite{Smirnov:2025mrm}.

The minimality has always been the guiding principle in making such a choice. For the leptonic sector, the simplest possibility among all the analysed models is the very first model discussed by Feruglio in \cite{Feruglio:2017spp}. It is based on the group $\Gamma_3 \simeq A_4$, which happens to be the smallest modular group with a 3-dimensional irreducible representation to which the three generations of leptonic doublets $L_i$ can be assigned. The neutrino masses are attributed in this model to some source of lepton number violation at a high scale $\Lambda_L$ and are described in terms of a dimension-5 Weinberg operator. The charged leptons obtain masses through an $A_4$ triplet flavon field $\phi$, while a triplet $Y_3^{(2)}$ with weight 2 completely determines the neutrino masses and leptonic mixing in terms of a single parameter $\tau$. This simple model correctly reproduces the solar angle and the ratio of the solar and atmospheric mass scales in accordance with the inverted ordering of neutrino masses for a specific choice of $\tau$. It, however, fails to obtain viable values of the two other angles, $\theta_{23}$ and $\theta_{13}$, consistent with the data. This led to investigations of more complex models involving different choices of representation assignments, field content, and/or symmetry group, leading to different classes of effective Yukawa couplings \cite{Penedo:2018nmg,Kobayashi:2018vbk,Criado:2018thu,Kobayashi:2018scp,Novichkov:2018yse,Novichkov:2018nkm,Novichkov:2018ovf,Ding:2019zxk,Kobayashi:2019mna,deMedeirosVarzielas:2019cyj,Ding:2019xna}. For example, an extensive analysis of possibilities within the $A_4$ group was performed in \cite{Ding:2019zxk}, considering both the Weinberg operator and seesaw mechanisms for neutrino mass generation. While none of the Weinberg-operator models can describe the data, several seesaw models can accommodate it, albeit with a more complicated Yukawa sector than originally envisaged in \cite{Feruglio:2017spp}.

Our aim in this paper is to reanalyse the aforementioned simplest model, listed as Example-1 by Feruglio in \cite{Feruglio:2017spp}. We find that the original model in its simplest form contains a hitherto unnoticed source of neutrino masses, whose inclusion allows for a very good fit to all leptonic observables. This new source arises from $R$-parity violation \cite{Aulakh:1982yn,Hall:1983id,Ross:1984yg,Joshipura:1994wm,Romao:1999up,Hirsch:2000ef,Joshipura:1998fn}, see \cite{Barbier:2004ez} for a review and references. If lepton number conservation is not insisted upon, then this term is allowed in the K\"ahler potential of the model without altering the field content or symmetry of the original model. This lepton-number-violating term generates an effective bilinear $R$-parity-violating (RPV) term $\epsilon_i L_i H_u$ in the superpotential once supersymmetry is broken in theories with local supersymmetry. The mechanism is identical to the Giudice--Masiero mechanism for generating the $\mu$ term \cite{Giudice:1988yz}. The $\epsilon_i$ are determined by the modular form $Y_3^{(2)}(\tau)$ and its derivative and introduce two additional parameters compared to the original model. Including this contribution allows for both normal and inverted neutrino mass orderings and describes the entire neutrino spectrum in terms of only three or four parameters, depending on the modulus contribution to supersymmetry breaking.

The same RPV term in the K\"ahler potential also allows the fermionic submultiplets of the flavon triplet chiral superfield $\phi$ to couple to $L_i$, generating a Dirac mass term. Thus, if massive, the $\phi$ fermions automatically play the role of right-handed neutrinos. This opens the possibility of a seesaw model within the same framework, leading to a setup similar to the model discussed as Example-2 by Feruglio in \cite{Feruglio:2017spp}. Unlike the latter, we show that the seesaw version in our framework, together with the RPV contribution, can precisely describe the leptonic spectrum and yield a phenomenologically consistent neutrino mass spectrum.

\section{Supergravity Realisation of Feruglio’s Minimal Model}
\label{sec:sugra}
We first reformulate Feruglio's original model in supergravity rather than global supersymmetry to implement the Giudice--Masiero (GM) mechanism for $R$-parity violation. This requires a modification of the relation between the modular weights of the matter fields relative to that adopted in \cite{Feruglio:2017spp}. Supersymmetry (SUSY) breaking is necessary for implementing the GM mechanism. We assume that one source of SUSY breaking can be the $F$-term of the modulus $\tau$, but do not consider it to be the sole source. An additional source of SUSY breaking is also present and contributes to the gravitino mass, $m_{3/2}$. It may arise, for example, from a hidden-sector superpotential, gaugino condensation or dilaton dynamics. The SUSY-breaking scales associated with these two sources need not be the same, and one can dominate over the other in realistic cases. As we will argue, this is required for the consistency of the model.

Recall \cite{Feruglio:2017spp} that the complex modulus $\tau$, with ${\rm Im}(\tau)>0$, and a chiral superfield $\Phi_I$ with weight $k_{\Phi_i}$ transforms as
\beqa \label{mod}
\tau &\rightarrow & \gamma \tau = \frac{a\tau + b}{c\tau + d}\,,\nonumber \\
\Phi_I  &\rightarrow &  (c \tau + d)^{-k_{\Phi_I}}\,\rho(\gamma)\,\Phi_I\,,\eeqa
under the transformations of the homogeneous modular group with $a,b,c,d \in {\mathbb Z}$ and $ad-bc=1$. Here, $\rho(\gamma)$ denotes the representation of the corresponding finite modular group $\Gamma_N$ according to which $\Phi_I$ transforms. For the model of our interest, $\Gamma_N \simeq A_4$, under which the leptonic superfield $L$ and a flavon $\phi$ transform as $3$-dimensional irreducible representations, while the three generations of the charged-lepton superfields $E^c_i$ transform as $1$, $1^{\prime\prime}$ and $1^\prime$, respectively. The Higgs superfields $H_{u,d}$ are ordinary singlets under $A_4$. The modular weights of $L$, $H_u$, $H_d$, $\phi$, $E_i^c$ are denoted by $k_L$, $k_u$, $k_d$, $k_\phi$ and $k_{E^c_i}$, respectively.

Generalisation of the original model to local SUSY implies that K\"ahler potential $K \equiv K(\tau,\taub;\Phi_I,\bar{\Phi}_I)$ and superpotential $W \equiv W(\tau, \Phi_I)$ cannot be treated as independent functions. The model is instead characterised by a combination,
\be \label{G}
{\cal G}=K + \ln W + \ln \overline{W},\ee
which is required to be invariant under the modular transformations. For a usual K\"ahler potential of the form
\be \label{K}
K = K_0(\tau,\taub) + \sum_I Z^{k_{\Phi_I}}\,|\Phi_I|^2 +...\,, \ee
where
\be \label{K0_Z}
K_0 = - \ln(-i\tau + i\taub)\,,~~Z = (-i\tau + i\taub)^{-1}\,.\ee
The transformation of $K$ is
\be \label{K_trans}
K\, \rightarrow\, K\,+\, \ln(c\tau+d)\,+\,\ln(c\taub+d)\,.\ee
This can be compensated by the second term in Eq. (\ref{G}) if $W$ transforms as
\be \label{W_trans}
W\,\rightarrow\, (c\tau+d)^{-1}\, W\,.\ee
Therefore, $W$ must carry total weight $+1$ in this generalisation to supergravity, provided $K$ has the simplest aforementioned form.

Let us now revisit Feruglio's model in light of this generalisation. The charged lepton masses come from 
\be \label{W_e}
W_e=\frac{1}{\Lambda}\left(\alpha\, [L\phi]_1\,E_1^c\,+\,\beta\, [L\phi]_{1^\prime}\, E_2^{ c}\,+\,\gamma\,[L\phi]_{1^{\prime\prime}}\, E_{3}^c\right)\,H_d\,,\ee
where $[...]_r$ denotes the representation $r$ of $A_4$ constructed from the tensor product of superfields within it. The vacuum expectation value (VEV) of the flavon is assumed to have the form
\be \label{phi_vev}
\frac{\vev{\phi}}{\Lambda}=\left(\frac{\vev{\phi_1}}{\Lambda},0,0 \right)^T \equiv \left(u,0,0\right)^T\,.\ee
This breaks $A_4$ to a $Z_3$ subgroup and leads to a diagonal mass matrix for the charged leptons. The neutrino mass originates from the Weinberg operator
\be\label{W_nu}
W_\nu=\frac{1}{\Lambda_L}[L L Y]_1\, H_u H_u\,,\ee
where $Y$ is a triplet containing three linearly independent weight 2 modular forms, $Y_i \equiv Y_i(\tau)$. This completely determines the structure of the neutrino mass matrix,
\be \label{Mn1}
M_\nu = \frac{v_u^2}{\Lambda_L} \left(\ba{ccc} 2Y_1 & -Y_3 & -Y_2 \\ -Y_3 & 2 Y_2 & -Y_1 \\ -Y_2 & -Y_1 & 2 Y_3 \ea \right)\,,\ee
leading to a very predictive framework for the neutrino masses and mixing. Along with this, the model also contains a usual $\mu$-term,  
\be \label{W_mu} 
W_\mu = - \mu H_u H_d\,,\ee
which does not play a direct role in determining the lepton masses and mixing in the $R$-parity-conserving case. However, it will become relevant in the case we describe next.

Generalisation to local SUSY and the resulting requirement, Eq. (\ref{W_trans}), for the above $W_e$, $W_\nu$ and $W_\mu$ lead to the following constraints on the combinations of the weights: 
\beqa \label{k_const}
k_{E_i^c}+k_d+k_L+k_\phi&=& 1\,,\nonumber \\
2(k_L+k_u)-2&=&1\,,\nonumber \\
k_d+k_u&=&1 \,. \eeqa
The second condition uniquely fixes the modular weight of $LH_u$ to $3/2$. As argued in \cite{Feruglio:2017spp}, other operators involving $\phi$, or combinations of $\phi$ and $Y$, must be forbidden in $W_\nu$ to maintain the predictivity of the simplest structure of $M_\nu$. This is ensured by requiring $k_\phi < -2$ in the present case. The negative modular weight of $\phi$ also guarantees that no higher-dimensional operators involving $\phi$ and $Y$ can appear in $W_e$.

\section{$R$-Parity Violation via Giudice–Masiero Mechanism}
\label{sec:GM}
The bilinear $R$-parity violation, if enabled, can provide an additional source of lepton number violation and neutrino masses \cite{Barbier:2004ez} within this setup. This would be desirable since the minimal model fails to reproduce the observed data. The usual bilinear RPV term in the superpotential is forbidden by holomorphicity if $k_\phi < -2$ within the present setup. However, one can introduce a term in the K\"ahler potential,
\be \label{KGM}
K_{\rm GM} = \frac{1}{\Lambda}\, Z^{-2}\,\left[Y^\dagger(\taub) \left(\phi L\right)\right]_1 H_u + {\rm h.c.}\,, \ee
provided $k_\phi = -7/2$. The combination of the superfields and $Y^\dagger(\taub)$ above transforms with a factor of $(c \tau+d)^2\,  (c \taub +d)^2$ which is compensated by the transformation of $Z^{-2}$, leaving the entire term invariant under the action of modular transformation.

Once the SUSY is broken by the $F$-terms of the moduli and/or hidden sector fields, it induces an effective trilinear term in the superpotential. Computation of this for sub-Planck scale $\Lambda$, following the generalisation of \cite{Giudice:1988yz,Kaplunovsky:1993rd}, leads to (see Appendix \ref{app:triliner_GM})
\be \label{W_eff}
W_{\rm eff} = \frac{1}{\Lambda} \left[ \left(m_{3/2} Z^{-2} Y - F^{\taub} \partial_{\taub} (Z^{-2} Y)\right)\, \left(\phi L\right)\right]_1 H_u \,. \ee
Further simplification allows one to express 
\be \label{W_eff_Dirac}
W_{\rm eff} = (Y_D)_{ij}\,\phi_i L_j H_u\,,\ee
in terms of the canonically normalised fields, where
\be \label{YD}
Y_D = -\frac{m_{3/2} g_1}{\Lambda} \left(\ba{ccc} 2 X_1 & (g-1) X_2 & -(g+1) X_3 \\ -(g+1) X_2 & 2 X_3 & (g-1) X_1 \\ (g-1) X_3 & -(g+1) X_1 & 2 X_2 \ea \right)\,,\ee
and $g=g_2/g_1$. Here, $g_{1,2}$ result from the fact that there are two independent contractions, i.e. $[Y^\dagger(\taub) (\phi L)_{3_S}]_1$ and $[Y^\dagger(\taub) (\phi L)_{3_A}]_1$ allowed in $K_{\rm GM}$. They are suitably rescaled to absorb the factors arising from the normalisation of kinetic terms. Also, 
\be \label{X_i}
X_i(\tau) = -Z^{-2}  Y_i^*(\tau) + \frac{\langle F^{\taub}\rangle}{m_{3/2}}\left(2i Z^{-1} Y_i^*(\tau)+Z^{-2}(\partial_\tau Y_i(\tau))^*\right)\,.\ee
Note that $(Y_D)_{ij}$ are no longer restricted to be holomorphic functions of modular forms, as they effectively originate from the K\"ahler potential through SUSY breaking.

Once the scalar component of $\phi_i$ acquires non-vanishing vacuum, see Eq. (\ref{phi_vev}),  $W_{\rm eff}$ leads to an effective bilinear RPV term parametrised as
\be \label{ep}
W_{\slashed{R}} = (Y_D)_{1i} \langle \phi_1 \rangle\, L_i H_u\ \equiv -\epsilon_i\, L_i H_u\,.\ee
As is well known \cite{Joshipura:2002fc}, the generation of $\epsilon_i$ is not sufficient to obtain neutrino masses, since the $\epsilon_i$ term can be rotated away from the superpotential by redefining $L_i$ and $H_d$. The $\epsilon_i$ cannot, however, be simultaneously removed from the soft SUSY-breaking terms if they are non-universal. This leads to sneutrino VEVs, and in turn, masses for the neutrinos.

Following the convention used in \cite{Joshipura:2021vtf}, we parameterize the relevant soft terms as,
\beqa \label{V_soft}
V_{\rm soft} &=& m_{L_i}^2\, |\tilde{L}_i|^2 + m_{H_d}^2\, |\tilde{H}_d|^2 + m_{H_u}^2\, |\tilde{H}_u|^2\, \nonumber \\
 &+& \left\{B_\mu\,\mu\, \tilde{H}_u \tilde{H}_d + A_i\,\epsilon_i\, \tilde{L}_i \tilde{H}_u + {\rm h.c.}\right\}\,.\eeqa
Computation of the above soft parameters for the K\"ahler potential given in Eq. (\ref{K}), following \cite{Kaplunovsky:1993rd}, leads to
\beqa \label{soft}
m_{\Phi_I}^2 &=& m_{3/2}^2 - k_{\Phi_I}\,|F^{\tau}|^2\,Z^2\,,\nonumber\\
A_i &=& - F^{\tau}\left(\partial_\tau \ln X_i - i Z \left(k_\phi + k_L+k_u-\frac{1}{2}\right)\right)\,,\nonumber\\
B_\mu &=& m_{3/2} + i F^{\tau}Z \left(k_u + k_d - \frac{1}{2} \right)\,.\eeqa
Consequently, a non-zero $F^\tau$ gives rise to non-universal soft SUSY-breaking parameters at the mediation scale itself in the present framework. The structure of the $F^{\tau}$-induced soft SUSY breaking and some of its phenomenological aspects have been discussed in \cite{Kobayashi:2021jqu,Tanimoto:2021ehw,Kikuchi:2022pkd,Ding:2022nzn}.

The sneutrino VEVs in the basis in which the bilinear terms are rotated away from $W$ can be expressed as \cite{Joshipura:2002fc} $\omega_i = \kappa_i \epsilon_i$, where
\be \label{ki}
\kappa_i \simeq \frac{v_d}{\mu}\,\frac{(m_{H_d}^2 - m_{L_i}^2) - \mu \tan\beta\, (B_\mu - A_i)}{m_{L_i}^2 + \frac{1}{2} M_Z^2 \cos 2\beta}\,, \ee
is obtained by minimizing the full scalar potential \cite{Giudice:1992jg,Joshipura:2021vtf,Joshipura:2023ewa}. The sneutrino VEVs mix neutrinos with neutralinos and generate a contribution to the neutrino masses when the neutralinos are integrated out. In the seesaw limit ($\omega_i \ll \mu, M_{1,2}$, where $M_{1,2}$ are the gaugino soft masses), this contribution is given by\footnote{In addition to this, $\epsilon_i$  also generates a loop-induced \cite{Barbier:2004ez} contribution to neutrino masses. This is sub-dominant compared to the tree-level contribution considered here, and will be neglected.}
\be \label{Mn2}
\left(M_ {\nu \slashed{R}}\right)_{ij} = A_0\,\omega_i \omega_j\,,\ee
where,
\be \label{A0}
A_0 = - \frac{\mu (g^2 M_1 + g^{\prime 2} M_2)}{2(M_1 M_2 \mu - v_u v_d (g^2 M_1 + g^{\prime 2} M_2))}\,.\ee

Before we reassess the phenomenological viability of the neutrino spectrum in light of this new source of neutrino masses, let us point out some features which crucially depend on whether $F^{\taub}$ is the only source of SUSY breaking. For convenience, we parametrize at the vacuum value of $\tau$,
\be \label{xi}
\frac{F^{\taub} Z}{m_{3/2}} \equiv \xi\,, 
\ee
where $|\xi| \leq \sqrt{3}$ in general, and the highest value is achieved if $F^{\taub}$ is the sole source of SUSY breaking\footnote{Assuming vanishing vacuum energy density for the supergravity scalar potential leads to a relation $m_{3/2}^2=\frac{1}{3} K_{I\bar{J}} F^I F^{\bar{J}}$, where $I, J$ stand for moduli and hidden sector fields. If only $F^{\taub}$ is nonvanishing, then $K$ specified in Eq. (\ref{K}) leads to $m_{3/2}^2=\frac{1}{3} |F^{\taub}|^2 Z^2$ at the minimum of the modulus.}. On the other hand, if $F^{\taub}$ is the only source of SUSY breaking, then the soft masses determined in Eq. (\ref{soft}), along with the relations in Eq. (\ref{k_const}) and $k_\phi=-7/2$, lead to the fact that $(2m_{L}^2 + m_{E^c}^2)/m_{3/2}^2 = 3-5\xi^2$. Requiring this quantity to be positive, to prevent the charge-breaking minima, leads to an even more stringent limit,
\be \label{xi_bound}
|\xi| \leq \sqrt{\frac{3}{5}}\,.\ee
Hence, $F^{\taub}$ cannot be the only source of SUSY breaking within the present setup.

Computation of $\kappa_i$ by substituting Eqs. (\ref{k_const},\ref{soft},\ref{xi}) in Eq. (\ref{ki}) leads to
\be \label{ki_2}
\kappa_i \simeq -\frac{v_u}{m_{3/2}} \frac{1}{(1-k_L \xi^2)}\left(1 + \frac{\xi}{Z} \partial_\tau \ln X_i + 3 i \xi - \frac{\xi^2 m_{3/2}}{2\mu \tan\beta}\right)\,, \ee
where we ignore ${\cal O}(M_Z^2/m_{3/2}^2)$ terms. Apparently, $\kappa_i$ do not vanish at high scale even in the $\xi \to 0$ limit. However, they are flavour universal in this limit. When $\xi \neq 0$, small but calculable non-universality among them gets induced through the second term in Eq. (\ref{ki_2}). Note that $\mu$ is an independent parameter in the present setup, and one expects $\mu \sim {\cal O}(m_{3/2})$ for low-energy SUSY. Alternatively, the $\mu$-term itself can be generated using the GM mechanism with a slight modification within this setup; see Appendix \ref{app:mu}. With $\mu \sim {\cal O}(m_{3/2})$ and neglecting ${\cal O}(\xi^2)$ terms, $\kappa_i$ at low energy can be written as $\kappa_i = \kappa_0 x_i$ with
\be \label{ki_3}
\kappa_0 = -\frac{v_u}{m_{3/2}}\,, ~~x_i \approx 1 + \frac{\xi}{Z}\, \partial_\tau \ln X_i  + 3 i \xi \,. \ee
An additional source of SUSY breaking in the conventional approach would introduce flavour-universal terms $m_0^2$, $m_{3/2}A_0$ and $m_{3/2}B_0$ in $m_{\Phi_I}^2$, $A_i$ and $B_\mu$, respectively. This replaces $1 \to 1+B_0-A_0$ in the first term of $\kappa_i$. For $F^{\taub}=0$, this coefficient can anyway be absorbed into the definition of $\kappa_0$. Otherwise, it constitutes an additional parameter. For definiteness, we set $A_0=B_0$ and neglect their contribution to $\kappa_i$ in the subsequent analysis.

\section{Neutrino Masses from Weinberg Operator and RPV}
\label{sec:model1}
Combining both the contributions, the neutrino mass matrix can be parameterised as
\be \label{Mnu}
M_\nu = \frac{v_u^2}{\Lambda_L}\,\left(\left(\ba{ccc} 2Y_1 & -Y_3 & -Y_2 \\ -Y_3 & 2 Y_2 & -Y_1 \\ -Y_2 & -Y_1 & 2 Y_3 \ea \right) + r\,{\cal X}\right)\,\ee
where
\be \label{X}
{\cal X} = \left(\ba{ccc} 4 \tilde{X}_1^2 & 2  (g-1)\tilde{X}_1 \tilde{X}_2 & -2 (g+1) \tilde{X}_1 \tilde{X}_3 \\ 2(g-1) \tilde{X}_1 \tilde{X}_2 & (g-1)^2 \tilde{X}_2^2 & (1-g^2) \tilde{X}_2 \tilde{X}_3\\
-2 (g+1)\tilde{X}_1 \tilde{X}_3 & (1-g^2) \tilde{X}_2 \tilde{X}_3 & (1+g)^2 \tilde{X}_3^2 \ea\right)\,,\ee
with $\tilde{X}_i = X_i\, x_i$ and $g = g_2/g_1$. $X_i$ involves terms anti-holomorphic in $Y_i$ as well as their derivatives. They are expressed more explicitly in Appendix \ref{app:modular}. The parameter 
\be \label{rr}
r = \frac{A_0 \kappa_0^2 u^2 m_{3/2}^2 g_1^2}{v_u^2/\Lambda_L}\,,\ee
quantifies the relative strength of two contributions. The leptonic mixing angles, CP phases and the ratio of solar to atmospheric squared-mass differences are completely determined by complex $\tau$, $g$ and real $r$, $\xi$.

\begin{table*}[t]
\centering
\small
\begin{tabular}{ccccc}
\hline
\hline
Parameters/ & dim-5 only & dim-5 $+$ RPV  & dim-5 $+$ RPV   & dim-5 $+$ RPV \\ 
Observables & (Ex. 1 in \cite{Feruglio:2017spp})  & (with $|F^{\taub}|=0$) & (with $|F^{\taub}| \neq 0$) &  (with $|F^{\taub}|\neq 0$) \\ 
\hline
Ordering & Inverted & Inverted & Inverted & Normal  \\
\hline
  $\tau$   &   $-0.01181 + 0.9947\,i$    &   $-0.00916 + 0.90881\,i$  & $0.43566 + 0.86613\,i$ & $-0.06592 + 0.99115\, i$\\
   $g$ &   -    &    $-19.8067 + 0.69569\,i$ & $-5.71644 - 11.1233\,i$ & $8.46215 - 1.91868\,i$\\
   $r$ &   -    &    $2.9728 \times 10^{-4}$ & $2.88203 \times 10^{-3}$ & $-0.888826 \times 10^{-3}$\\
   $\xi$ &   -    &  -  & $0.56981$ & $-0.41659$ \\
      \hline
    $\frac{\Delta m_{21}^2}{\Delta m_{3l}^2}$ &   $-0.0301$    &  $-0.0301$  & $-0.0301$ & $0.0299$\\
   $ \sin^2\theta_{12}$ &  $0.3086$     &    $0.3087$  &  $0.3091$ & $0.3088$\\
   $ \sin^2\theta_{23}$ &  $0.349$    &   $0.545$  & $0.556$ & $0.47$\\
   $ \sin^2\theta_{13}$ &  $0.04468$    &  $0.02262$   & $0.02261$ & $0.02249$\\
      \hline
   $\chi^2_{\rm min}$ & $1720$ & $0$ & $0$ & $0$ \\
     \hline
   $ \delta/\pi$ & $0.455$     &   $0.428$  & $-0.479$ & $-0.342$\\
   $ \alpha_{21}/\pi$ &  $0.214$     &    $0.306$  & $0.245$ & $0.216$\\
   $ \alpha_{31}/\pi$ &   $1.216$    &   $0.936$   &  $-0.878$ & $-0.301$\\  
   $ |m_{\beta \beta}|$ [eV] &  $0.045$ & $0.047$  & $0.046$  & $0.023$ \\
   $ |m_{\beta}|$ [eV] &  $0.048$     &    $0.052$  & $0.05$ & $0.026$\\
   $ \sum m_\nu$ [eV] &  $0.1$    &   $0.125$   & $0.108$ & $0.107$\\
      \hline
     \hline
\end{tabular}
\caption{Sample best-fit solutions for neutrino masses originating from the Weinberg operator, with and without $R$-parity violation, are parametrised in Eq. (\ref{Mn2}). The neutrino oscillation data used in the fits are taken from NuFIT 6.1 (2025) \cite{Esteban:2024eli} (IC23 without Super-Kamiokande atmospheric data). The $\chi^2$ function includes the ratio $\Delta m_{21}^2/\Delta m_{3l}^2$, where $l=1$ ($2$) for normal (inverted) ordering, and the three neutrino mixing angles. Predictions are given for the three CP-violating phases, defined according to the PDG convention \cite{ParticleDataGroup:2024cfk}, the effective mass governing neutrinoless double-beta decay, $|m_{\beta\beta}|$, the effective mass in ordinary beta decay, $|m_\beta|$, and the sum of the three neutrino masses, $\sum m_\nu$. The latter are obtained by normalising the fitted atmospheric mass scale to its corresponding global best-fit value.}
\label{tab:res1}
\end{table*}
Numerically, the non-vanishing RPV contribution significantly improves the fit to the current best-fit neutrino oscillation data, as shown in Table \ref{tab:res1}. We also compare with Feruglio’s original model, which was found to be closer to the data only for the inverted ordering. Remarkably, even for $|F^{\bar\tau}|=0$, the RPV contribution renders the model viable, although the resulting $\sum m_\nu$ is slightly above the current limit, $\sum m_\nu<0.12$ eV, from cosmological simulations \cite{Shao:2024mag}. This can be brought within the bound by allowing a small $|F^{\bar\tau}|$, in which case the $\kappa_i$ become non-universal. Interestingly, viable normal ordering is also obtained for $|F^{\bar\tau}|\neq0$.  \emph{Although the RPV contribution introduces additional parameters that may seemingly improve the fit trivially, they arise intrinsically within the framework, rather than from any model extension or enlarged symmetry.}

Using the fitted value of $r$ in Table \ref{tab:res1} and Eqs. (\ref{rr},\ref{ki_3}) along with $A_0 \sim m_{3/2}^{-1}$, $u \sim 10^{-2}$, we find $g_1^2 \sim m_{3/2}/\Lambda_L$.  Note that $g_1$ sets the overall strength of the term in $K_{\rm GM}$, and for values close to the above, the RPV contribution plays a non-trivial and decisive role in making the minimal model consistent with the data. It is to be noted that the required strength of the $R$-violating contribution compared to the Weinberg operator as measured by $r$ is less than $0.3\%$.

\section{Seesaw Version}
\label{sec:model2}
The $W_{\rm eff}$ in Eq. (\ref{W_eff_Dirac}) induced by the non-minimal K\"ahler term also gives rise to a Dirac-type Yukawa interaction term between the $L_i$ and the fermionic components of $\phi_i$. If the latter acquire large but finite masses, then one naturally obtains a seesaw contribution to the neutrino masses within this setup. Essentially, the fermionic partners in $\phi_i$ play the role of right-handed neutrinos. Moreover, the VEVs of their scalar counterparts break $R$-parity spontaneously, leading to a framework like that in \cite{Chun:1995js,Chun:1995bb}, studied earlier but without modular symmetry.

To generate masses for $\phi_i$ fermions, we consider the simplest possibility of introducing a bare mass term, $\frac{M_T}{2}(\phi \phi)_1$. This leads to the following mass matrix for the fermion triplet:
\be \label{Mphi}
M_\phi = M_T \left(\ba{ccc} 1 & 0 & 0 \\ 0 & 0 & 1 \\ 0 & 1 & 0 \ea \right)\,.\ee
The bare mass term sets $k_\phi = 1/2$. This along with the $K_{\rm GM}$, $W_e$ and $W_\mu$ implies $k_u = -5/2-k_L$, $k_d=7/2+k_L$ and $k_{E^c_i}=-3-2k_L$. Consequently, $(L H_u)^2$ carries a weight $-5$ and $\phi$ being the only flavon with positive weight, the lowest-order Weinberg operator arises only at $(\phi/\Lambda_L)^{12}$, making it vanishingly small for $\Lambda_L \geq \Lambda$. Similarly, the lowest order correction to $W_e$ arises at $(\phi^4 Y/\Lambda^4)$ and it is also negligible for $u \sim 10^{-2}$.  

For $M_T \gg v_u$, the seesaw-induced light neutrino mass matrix is given by
\be \label{Mnu_ss0}
 - v_u^2\, Y_D^T\,M_\phi^{-1}\,Y_D \equiv \frac{v_u^2}{M_T} \frac{m_{3/2}^2 g_1^2}{\Lambda^2}\, {\cal S}\,, \ee
where 
\beqa \label{}
{\cal S}_{11} &=& 2 \left(g^2-1\right) X_2 X_3-4 X_1^2\,, \nonumber \\
{\cal S}_{12} &=&  -\left(g^2+4 g-1\right) X_1 X_2-2 (g-1) X_3^2\,, \nonumber \\
{\cal S}_{13} &=&  \left(-g^2+4 g+1\right) X_1 X_3+2 (g+1) X_2^2\,, \nonumber \\
{\cal S}_{22} &=& 4 (g+1) X_1 X_3-(g-1)^2 X_2^2\,, \nonumber \\
{\cal S}_{23} &=&  \left(g^2-1\right) X_1^2+\left(g^2-5\right) X_2 X_3\,, \nonumber \\
{\cal S}_{33} &=&  -(g+1)^2 X_3^2-4 (g-1) X_1 X_2\,.\eeqa
This, along with the RPV contribution of Eq. (\ref{Mn2}), can be combined to write the full neutrino mass matrix as
\be \label{Mnu_ss}
M_\nu =\ \frac{v_u^2}{M_T} \frac{m_{3/2}^2 g_1^2}{\Lambda^2}\,\left({\cal S} + r\,{\cal X}\right)\,,\ee
where $r$ in the present case is given by
\be \label{rr_ss}
r = \frac{A_0 \kappa_0^2 u^2 \Lambda^2}{v_u^2/M_T}\,.\ee
Comparing with the previous case, this essentially implies that $M_T$ is a source of lepton-number violation, with $M_T \sim (g_1 m_{3/2}/\Lambda)^2 \Lambda_L$. However, the flavour structures of the mass matrices are very different.

\begin{table}[t]
\centering
\small
\begin{tabular}{ccc}
\hline
\hline
Parameters/ & Seesaw only & Seesaw $+$ RPV  \\ 
Observables &   & (with $|F^{\taub}|=0$)\\ 
\hline
Ordering & Normal & Normal \\
\hline
  $\tau$   &   $0.00778 + 1.00234\,i$    &   $0.04137 + 1.00647\,i$  \\
   $g$ &   $-3.62656 - 0.30739\, i$    &    $-4.08806 - 0.02553\,i$\\
   $r$ &   -    &    $-0.27744$ \\
      \hline
    $\frac{\Delta m_{21}^2}{\Delta m_{3l}^2}$ &   $0.0299$    &  $0.0299$  \\
   $ \sin^2\theta_{12}$ &  $0.309$     &    $0.3088$  \\
   $ \sin^2\theta_{23}$ &  $0.395$    &   $0.47$\\
   $ \sin^2\theta_{13}$ &  $0.02235$    &  $0.02249$ \\
      \hline
   $\chi^2_{\rm min}$ & $20$ & $0$ \\
     \hline
   $ \delta/\pi$ & $0.031$     &   $-0.174$\\
   $ \alpha_{21}/\pi$ &  $0.432$     &    $-0.01$\\
   $ \alpha_{31}/\pi$ &   $-0.172$    &   $-0.56$\\  
   $ |m_{\beta \beta}|$ [eV] &  $0.002$ & $0.002$  \\
   $ |m_{\beta}|$ [eV] &  $0.009$     &    $0.009$\\
   $ \sum m_\nu$ [eV] &  $0.059$    &   $0.06$\\
      \hline
     \hline
\end{tabular}
\caption{Same as Table \ref{tab:res1}, but for the neutrino masses arising from the seesaw mechanism with/without $R$-parity violation, Eq. (\ref{Mnu_ss}).}
\label{tab:res2}
\end{table}
Numerically, this case provides an excellent fit to the data, as shown in Table \ref{tab:res2}. Even in the limit of vanishing RPV contribution, a good fit is obtained for normal ordering, with only $\sin^2\theta_{23}$ deviating by more than $3\sigma$ from its experimental value. Switching on the RPV contribution, even with $|F^{\taub}|=0$, yields an excellent fit to the data. An equally good fit is also obtained for inverted ordering. The comprehensive predictions for both orderings are given in Appendix \ref{app:pred}.

The values of $r$ extracted from the numerical solutions in Table \ref{tab:res2}, along with $A_0 \sim m_{3/2}^{-1}$, $u \sim 10^{-2}$ in Eq. (\ref{rr_ss}) leads to
\be \label{Lambda_ss}
\Lambda^2 \sim 10^3\, \frac{m_{3/2}^3}{M_T}\,. \ee
Hence, $\Lambda$ cannot be made arbitrarily large in the present case. Also, $M_T$ cannot be much larger than $m_{3/2}$ without lowering $\Lambda$. One finds $\Lambda \sim {\cal O}(10\, m_{3/2})$ for $M_T \sim m_{3/2}$. Recall that this constraint follows from the requirement that both the contributions to neutrino masses are comparable in size. Moreover, reproducing the scale of the atmospheric squared-mass difference from the fits requires the overall coefficient in Eq. (\ref{Mnu_ss}) to be of ${\cal O}(10^{-13})$ GeV. This, along with Eq. (\ref{Lambda_ss}), implies
\be \label{g1_2}
g_1^2 \sim 10^{-10}\times \left(\frac{m_{3/2}}{10\,{\rm TeV}}\right)\,.\ee
Nevertheless, the small value of $g_1$ is technically natural.

\section{Summary}
\label{sec:summary}
We have reanalysed the basic idea of describing neutrino masses solely in terms of modular forms. Taking the minimal modular $A_4$ model proposed by Feruglio as an example,  we have demonstrated that supersymmetric models based on modular symmetry admit an additional source of neutrino masses through $R$-parity violation. The latter is inherently present in these models once lepton number conservation is not insisted upon. RPV terms may be forbidden in the superpotential by the choice of symmetry and field content, but they can arise in the K\"ahler potential and, within the framework of local supersymmetry, reappear in the effective superpotential after SUSY breaking. The resulting RPV contribution to the neutrino mass matrix is also described by the modular forms, but this is not necessarily restricted to holomorphic functions. We have shown that this additional contribution can provide an important, and in some cases crucial, correction to the neutrino mass matrix, qualitatively altering the viability of the model. In particular, we considered examples of the simplest models proposed by Feruglio based on $A_4$ modular symmetry, which are not in precise agreement with the neutrino oscillation data. Including the RPV contribution restores their viability and yields an excellent fit to the data.

\begin{acknowledgements}
This research at the Physical Research Laboratory was supported by the Department of Space (DOS), Government of India. We acknowledge the use of Grammarly for language corrections to improve manuscript readability. No other AI tools were used, and no scientific concepts, data, result and text were generated by AI.
\end{acknowledgements}

\appendix
\section{Supergravity-induced effective trilinear term}
\label{app:triliner_GM}
This Appendix aims to extend the Giudice-Masiero (GM) mechanism \cite{Giudice:1988yz} to include a non-renormalisable trilinear term in the K\"ahler potential. We closely follow the notations and conventions used in \cite{Kaplunovsky:1993rd}, which also reproduces the standard GM result.

The K\"ahler potential of our interest is
\be \label{K2}
K = M_P^2\, K_0(\tau,\taub) + Z_I(\tau,\taub)\,\bar{Q}^{\bar I} Q_I + \left(\frac{1}{6 \Lambda} H_{IJK}(\tau,\taub)\, Q_I Q_J Q_K+{\rm h.c.} \right) \,.\ee
Here, $Q_I$ are visible sector fields. The dimensionless functions $K_0$, $Z_I$ and $H_{IJK}$ can also include hidden-sector fields and/or dilaton in addition to moduli in the more general considerations. Consider the superpotential with a tree-level non-zero trilinear term,
\be \label{W2}
W =  \frac{1}{3} \tilde{Y}_{IJK}\, Q_I Q_J Q_K\,,\ee
for generality although in our setup we have $\tilde{Y}_{IJK}=0$ for the fields of our interest.

The effective coefficient of bilinear term, involving fermions $\chi_I$ of the respective superfields $Q_I$, in the expansion of Lagrangian is given by \cite{Wess:1992cp}
\be \label{bl}
{\cal L} \supset -\frac{1}{2} e^{K/(2 M_P^2)}\, {\cal D}_I D_J W\, \chi_I \chi_J\,.\ee
Here, 
\beqa \label{DDW}
D_I W &=& \partial_I W + \frac{1}{M_P^2} (\partial_I K)\, W\,, \nonumber \\
{\cal D}_I D_J W &=& W_{IJ} + \frac{K_{IJ}}{M_P^2} W + \frac{K_I}{M_P^2} D_J W + \frac{K_J}{M_P^2} D_I W - \frac{K_I K_J}{M_P^4} W - \Gamma_{IJ}^K D_K W\,, \eeqa
where $W_{IJ} = \partial_I \partial_J W$ and so on. In the present case, the indices $I, J$ stand for the visible sector field, while $K$ also includes the hidden sector fields, including moduli. To evaluate the effective trilinear term, we consider the coefficient $e^{K/(2 M_P^2)} {\cal D}_I D_J W$ in Eq. (\ref{bl}) as a container of a scalar multiplet, say $\phi_K$, which then effectively induces a trilinear term. Therefore, the coefficient of $\phi_K \chi_I \chi_J$ can be extracted using
\be \label{Y0}
\langle \partial_K (e^{K/(2M_P^2)}\,{\cal D}_I D_J W) \rangle \equiv Y_{IJK}\,.\ee
This is the defining relation for extracting the effective trilinear coupling in the superpotential. Note that $Y_{IJK}$ are not yet canonically normalised.

Noting that $Q_I = 0$ for all the visible sector fields and $W=W_0$ at the minimum, the above can be simplified to  
\be \label{Y1}
Y_{IJK} = e^{K_0/2}\, \left(\tilde{Y}_{IJK} + \frac{\langle K_{IJK} \rangle}{M_P^2} W_0 -  \langle (\partial_K \Gamma_{IJ}^\tau)\, D_\tau W \rangle \right)\,.\ee
Further computation using $\Gamma^\tau_{IJ} = g^{\tau \bar{K}}\,\partial_I \partial_J \partial_{\bar{K}} K$, non-vanishing $F^{\taub}$ as \cite{Kaplunovsky:1993rd},
\beqa \label{Ftau}
F^{\taub} = \frac{1}{M_P^2} e^{K_0/2} (K_{0,\tau \taub})^{-1}\,D_\tau W\,,\eeqa 
and the fact that $e^{K_0/2} W_0/M_P^2 \equiv m_{3/2}$ implies 
\be \label{Y2}
Y_{IJK} = e^{K_0/2}\, \tilde{Y}_{IJK} + \frac{1}{\Lambda} \left(m_{3/2} H_{IJK} - F^{\taub} \partial_{\taub} H_{IJK} \right)\,.  \ee
This is the main result used in section \ref{sec:GM}.

\section{Generating $\mu$-term using GM mechanism}
\label{app:mu}
The $\mu$-term can also be induced within the setup discussed in section \ref{sec:model1} following the original proposal of Giudice-Masiero \cite{Giudice:1988yz}. One can write
\be \label{Kmu}
K_{{\rm GM}-\mu}=\frac{1}{\Lambda}\, Z^{-2}(\tau,\taub)\,[Y^\dagger(\taub) \phi]_1\,H_u H_d+ {\rm h.c.}\,, \ee
provided the last condition in Eq. (\ref{k_const}) is replaced by $k_d+k_u+k_\phi = -2$. In conjunction with the RPV term, these constraints lead to
\be \label{k_const_mu}
k_d=k_L,\, k_u = \frac{3}{2}-k_L,\, k_{E_i}^c=\frac{9}{2}-2k_L\,.\ee
Analogous to Eq. (\ref{xi_bound}), this case leads to a slightly relaxed limit $|\xi| \leq \sqrt{2/3}$.

The induced $\mu$ from $K_{{\rm GM}-\mu}$ is given by
\be \label{muterm}
\mu = -h\, m_{3/2} u\,X_1(\tau)\,,\ee
where $h$ is a dimensionless coefficient of ${\cal O}(1)$. Hence, $\mu$ becomes $\tau$-dependent. The corresponding soft term, in the convention $A_\mu \mu \tilde{H}_u \tilde{H}_d$, is given by
\beqa \label{Bmu}
A_\mu &=& - F^{\tau}\left(\partial_\tau \ln\mu - iZ \left(k_\phi+ k_d+k_u-\frac{1}{2} \right)\right)\,.\eeqa
Eventually, $\kappa_1$ vanishes at the high scale even when $\xi \neq 0$. The remaining ones are non-zero and can take distinct values. The parameter $A_0$ also becomes $\tau$-dependent through $\mu$.

In the limit $\xi \to 0$, all the $\kappa_i$ vanish at the high scale. They would be radiatively generated at low scale through the $b$ and $\tau$ Yukawa couplings \cite{Joshipura:2023ewa}. In this case, the induced values are small and universal, leading to $\kappa_i \simeq 10^{-3}$--$10^{-4}$. Other than this, the flavour structure of the neutrino mass matrices does not depend on whether $\mu$ is a bare term or generated by the above mechanism. Hence, all the results obtained with $|F^{\taub}| \to 0$ in Table \ref{tab:res1} and \ref{tab:res2} remain valid in this case too. This provides a quite elegant and minimal model in which the $\mu$-problem is addressed along with realistic neutrino masses and mixing.

\section{Modular forms and their derivatives}
\label{app:modular}
The modular forms of weight 2 for the underlying group can be written in terms of the Dedekind-eta function defined in the upper half-plane of the complex plane \cite{Feruglio:2017spp}, and can be expanded in terms of $q=\exp(2 \pi i \tau)$ as follows:
\beqa \label{Y_i}
Y_1(\tau) &=& 1 + 12 q + 36 q^2 + 12 q^3\,, \nonumber \\
Y_2(\tau) &=& -6 q^{1/3} \left(1 + 7 q + 8 q^2\right)\,, \nonumber \\
Y_3(\tau) &=& -18 q^{2/3} \left(1 + 2 q + 5 q^2\right)\,. \eeqa
Since ${\rm Im}(\tau) > 0$ leading to $|q|<1$, the expansion converges. For the faster convergence, we restrict ${\rm Im}(\tau)$ in the fundamental region, i.e. ${\rm Im}(\tau)> \sqrt{3}/2$, leading to $|q| < 0.0043$. 

The functions $X_i(\tau)$ in Eq. (\ref{X_i}) with the help of Eq. (\ref{xi}) can be expressed in a more compact form as, 
\be \label{X_i_2}
X_i(\tau) = -Z^{-2}\,\left((1-2 i \xi) Y_i^*(\tau) - \xi Z^{-1} \left(\partial_\tau Y_i(\tau)\right)^*\right)\,.\ee
Here, we have used $Y^\dagger(\taub) = Y^*(\tau)$ and $\partial_{\taub}Y^\dagger(\taub) = (\partial_\tau Y(\tau))^*$. The derivatives of $Y_i(\tau)$ are evaluated from Eq. (\ref{Y_i}) as,
\beqa \label{dY_i}
\partial_\tau Y_1(\tau) &=& 24 \pi i\, q \left(1 + 6 q + 3 q^2\right)\,, \nonumber \\
\partial_\tau Y_2(\tau) &=& -4 \pi i\,q^{1/3} \left(1 + 28 q + 56 q^2\right)\,, \nonumber \\
\partial_\tau Y_3(\tau) &=& -24 \pi i\, q^{2/3} \left(1 + 5 q + 20 q^2\right)\,.\eeqa
The derivative of $X_i(\tau)$ which determine the $\kappa_i$ in Eq. (\ref{ki_2}), evaluated using Eq. (\ref{X_i}) and reexpressed in terms of $\xi$ is given by,
\be \label{dX_i}
\partial_\tau X_i(\tau) = -2i Z \left(X_i(\tau) -  i \xi Z^{-2}\,  Y_i^*(\tau)\right)\,.\ee

\section{Comprehensive predictions for Seesaw $+$ RPV case}
\label{app:pred}
\begin{figure*}[t!]
\centering
\subfigure{\includegraphics[width=0.42\textwidth]{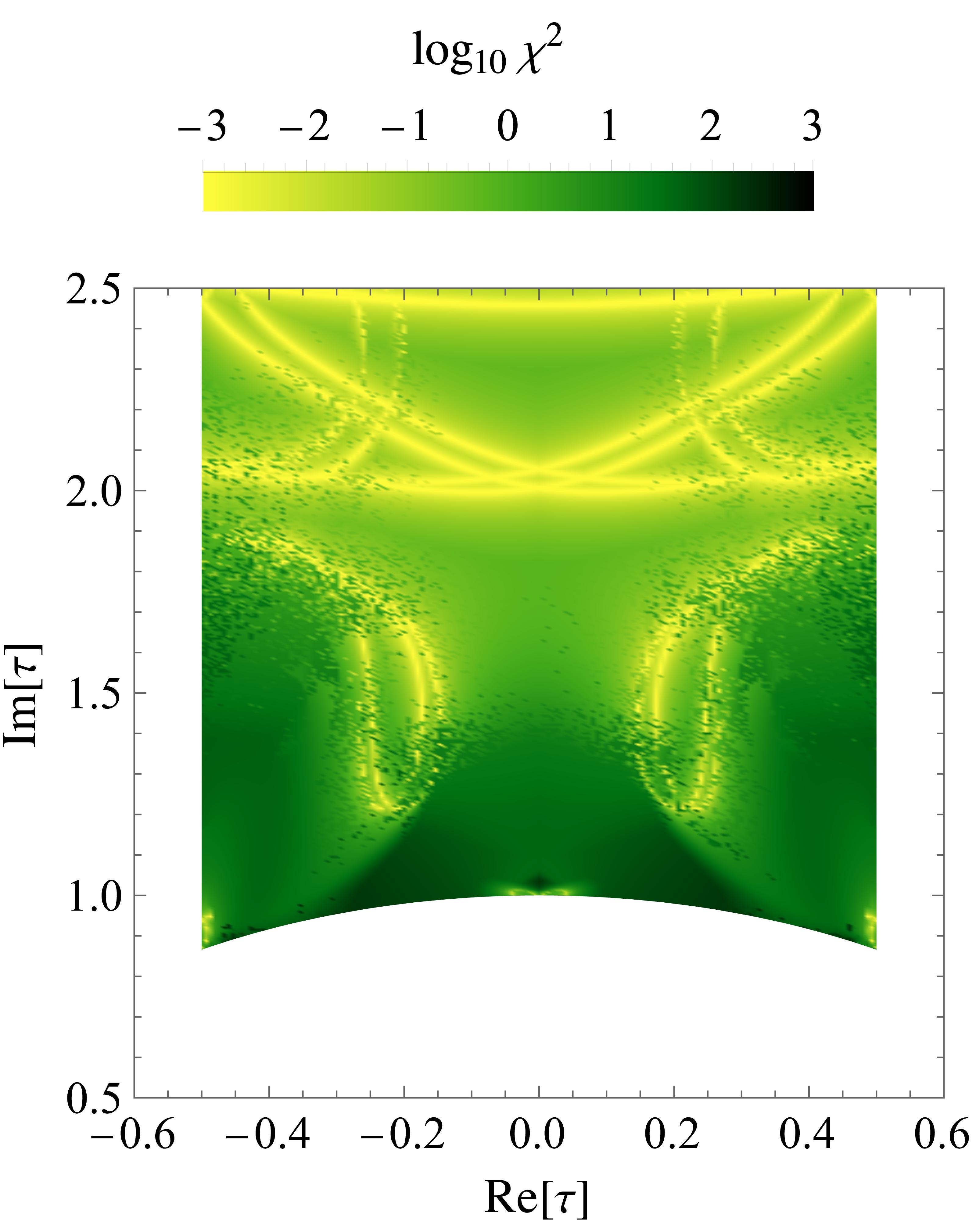}}\hspace*{0.5cm}
\subfigure{\includegraphics[width=0.42\textwidth]{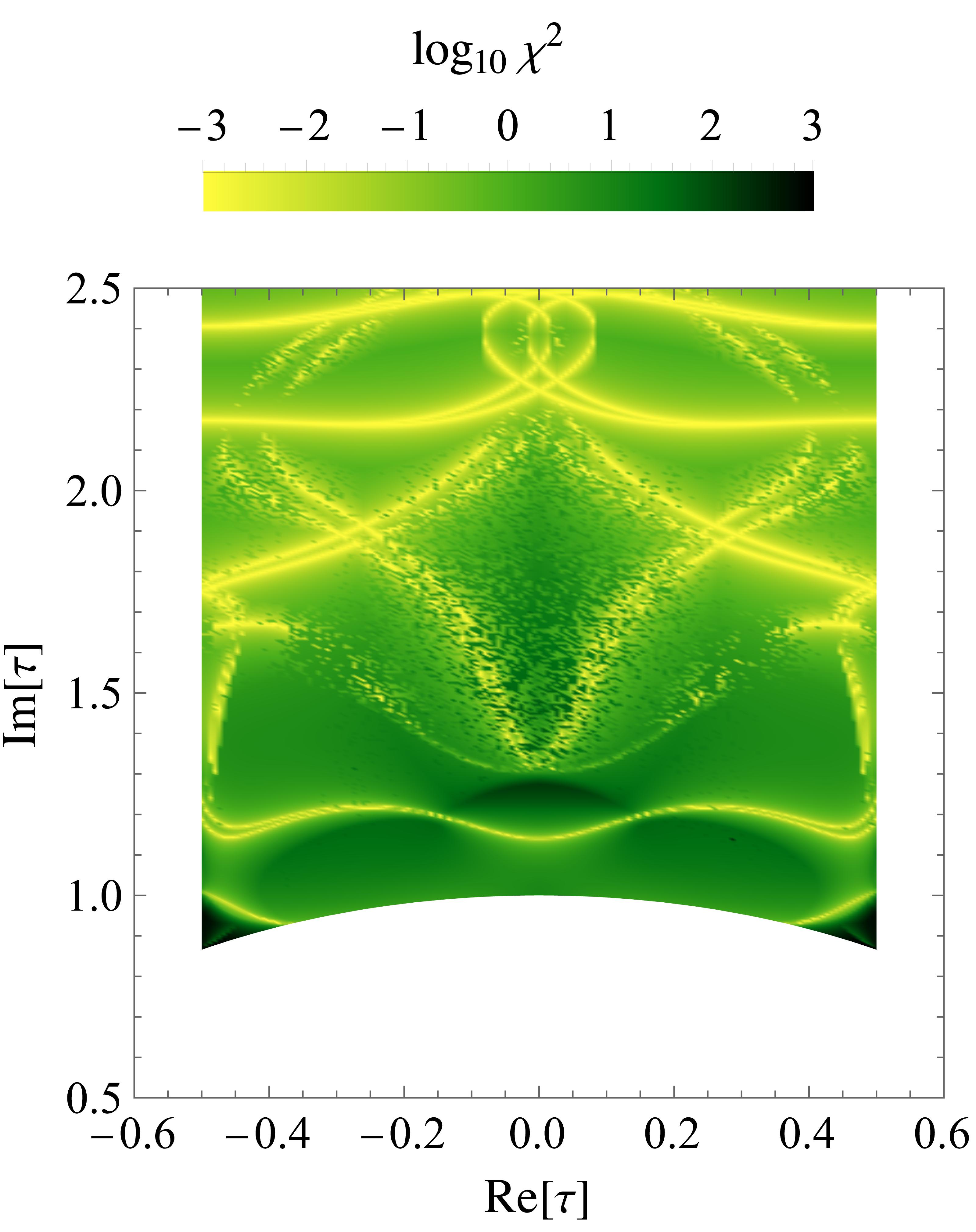}}
\caption{Minimised $\chi^2$ as a function of $\tau$ for the seesaw $+$ RPV case with $|F^{\taub}|=0$, assuming normal (left panel) and inverted (right panel) neutrino mass ordering. The $\chi^2$ function includes the ratio of the solar to atmospheric mass-squared differences and the three neutrino mixing angles, with the corresponding experimental values taken from NuFIT 6.1 (2025) \cite{Esteban:2024eli} (IC23 without Super-Kamiokande atmospheric data).}
\label{fig1}
\end{figure*}

As pointed out in section \ref{sec:model2}, the seesaw case along with RPV contribution to the neutrino masses characterised by $M_\nu$ in Eq. (\ref{Mnu_ss}) gives an excellent fit to the data both in the case of normal and inverted ordering. In this Appendix, we scan the $\tau$ minima in the fundamental region to find viability with the data. Assuming $|F^{\taub}|=0$, the values of the remaining parameters, $g$ and $r$, are optimised through $\chi^2$ minimisation. The results are displayed in Fig. \ref{fig1}. It is seen that this case provides a large set of $\tau$ values for which $\chi^2 < 1$ is obtained. This is valid in the case of both the normal and inverted orderings. 

\begin{figure*}[t!]
\centering
\subfigure{\includegraphics[width=0.42\textwidth]{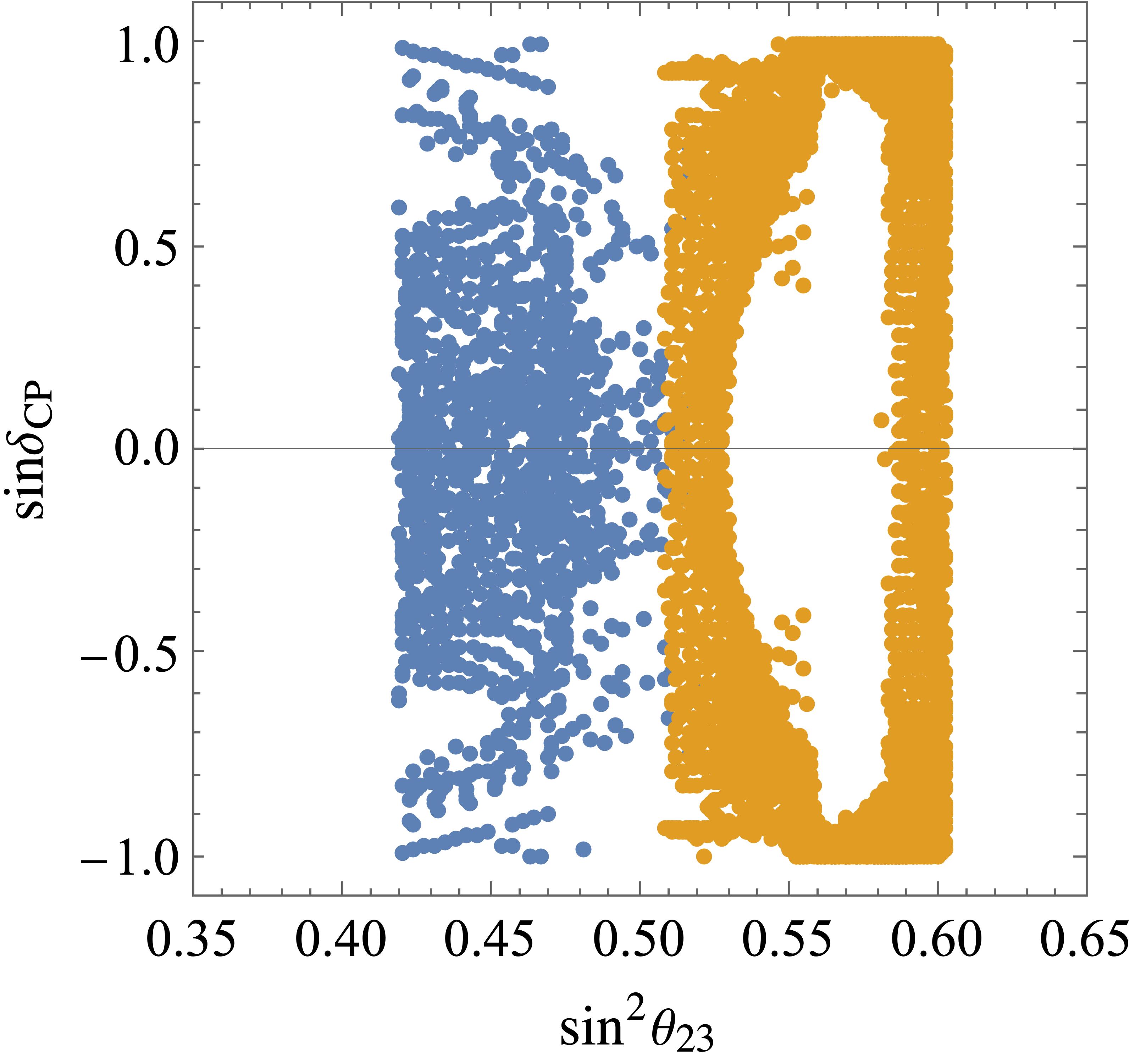}}\hspace*{0.5cm}
\subfigure{\includegraphics[width=0.42\textwidth]{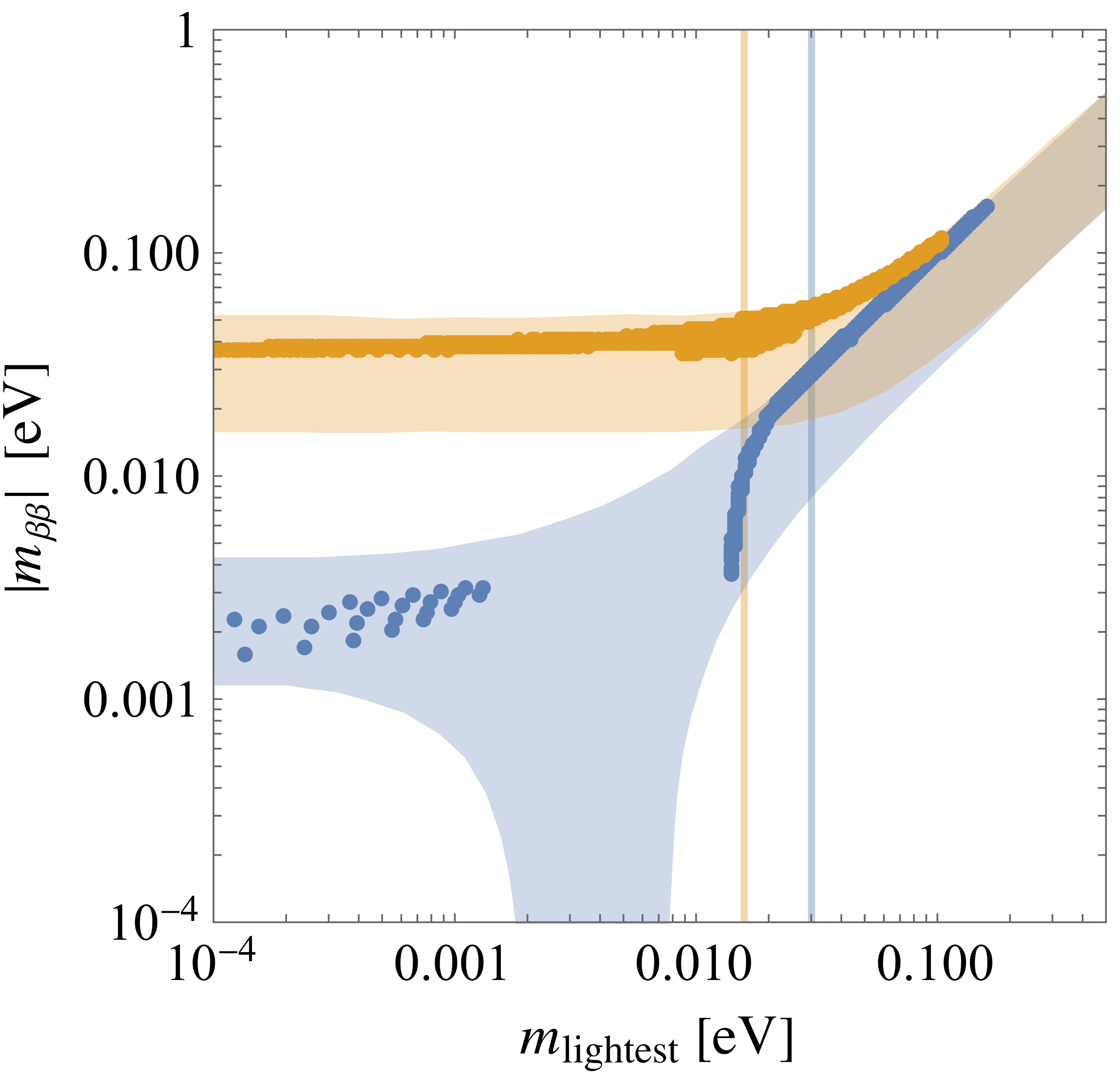}}
\caption{Predictions for various observables obtained from all best-fit solutions with $\chi^2 < 9$, for both normal (blue points) and inverted (orange points) mass ordering. In the left panel, only those points are shown which satisfy $\sum m_\nu < 0.12$ eV. The shaded region in the right panel indicates the most general region allowed at the $95 \%$ confidence level, without assuming any correlations among the observables. The vertical lines denote the upper limit on the lightest neutrino mass from cosmological observations.}
\label{fig2}
\end{figure*}

We collect all the solutions for which $\chi^2 < 9$ and study their corresponding predictions for the observables of current interest. This threshold ensures that none of the observables considered in the definition of $\chi^2$ goes beyond its $3\sigma$ range. The results obtained for the correlation between $\sin\delta_{\rm CP}$ and the atmospheric angle, as well as predictions of the effective mass of neutrinoless double beta decay, are shown in Fig. \ref{fig2}. In the case of the first, only those points are shown which satisfy the cosmological bound. It can be seen that the model allows almost all values of $\delta_{\rm CP}$. However, it has very specific predictions for the lifetime of neutrinoless double beta decay. 

Some of the best-fit points obtained, including the sample solutions shown in Table \ref{tab:res2}, lie very close to the fixed point $\tau=i$. It is well known that, exactly at this fixed point, the modulus leaves an unbroken $Z_2$ subgroup of $\Gamma_N$ \cite{Feruglio:2022koo,Feruglio:2023mii}. The generator of this $Z_2$ in the case of $\Gamma_N \simeq A_4$ is given by
\be \label{S}
\rho(S) = \frac{1}{3} \left(\ba{ccc} -1 & 2 & 2 \\ 2 & -1 & 2\\ 2 & 2 & -1 \ea \right)\,.\ee
At the fixed point, $Y_i(\tau)$ are restricted to satisfy \cite{Kashav:2024lkr}
\be \label{Yi_res}
Y_i(\tau=i) = (-i)^2\,\rho(S)_{ij}\, Y_j(\tau=i)\,,\ee
which leads to $\sum_i Y_i|_{\tau=i}= 0$. When $F^{\taub}=0$, this condition translates into $\sum_i X_i|_{\tau=i}= 0$, as can be seen from Eq. (\ref{X_i}). As a result, the seesaw-induced contribution characterised by ${\cal S}$ satisfies the invariant condition $\rho(S)^T\,{\cal S}\, \rho(S) = {\cal S}$. This, in turn, leads to a fixed-column prediction, $\frac{1}{\sqrt{3}}(1,1,1)^T$, in the neutrino mixing matrix.

In contrast to ${\cal S}$, the RPV contribution does not respect the same residual symmetry at $\tau=i$. In addition to the modulus, its origin also involves a flavon $\phi$ whose VEV does not respect the same $Z_2$ symmetry. The setup, therefore, provides an interesting example in which part of the contribution to the neutrino masses possesses a residual symmetry, while the other part can provide the required correction.

Restricting to $\tau=i$, the $M_\nu$ in Eq. (\ref{Mnu_ss}) along with $F^{\taub}=0$ leads to $\chi^2_{\rm min} = 11$. The solar angle is found to be $3.2\sigma$ away from its central value, while the remaining observables stay within $1 \sigma$. In the limit $r \to 0$, one finds a fixed-column prediction as mentioned above. Indeed, in this limit, we find $\Delta m_{21}^2/\Delta m_{31}^2 = 0.0695$, $\sin^2\theta_{12} = 0.3373$, $\sin^2\theta_{23} = 0.425$ and $\sin^2\theta_{13} = 0.01175$  which is qualitatively close to the pattern seen in the data. When the RPV contribution is added, it leads to a more accurate description.

\bibliography{ref}
\end{document}